\documentclass[twocolumn]{aastex63}

\usepackage{enumitem}
\usepackage{subcaption}
\submitjournal{Astrophysical Journal Supplement}

\graphicspath{{./}{figures/}}

\begin{document}

\title{Jupiter and Saturn as Spectral Analogs for Extrasolar Gas Giants and Brown Dwarfs}

\correspondingauthor{Daniel J. Coulter}
\email{dcoulter1331@gmail.com}

\author[0000-0001-7691-5096]{Daniel J. Coulter}
\affiliation{Department of Physics, University of Idaho, Moscow, ID 83844, USA}

\author[0000-0002-7755-3530]{Jason W. Barnes}
\affiliation{Department of Physics, University of Idaho, Moscow, ID 83844, USA}

\author[0000-0002-9843-4354]{Jonathan J. Fortney}
\affiliation{Department of Astronomy \& Astrophysics, University of California Santa Cruz, Santa Cruz, CA 95064, USA}




\begin{abstract}

With the advent of direct imaging spectroscopy, the number of spectra from brown dwarfs and extrasolar gas giants is growing rapidly. Many brown dwarfs and extrasolar gas giants exhibit spectroscopic and photometric variability, which is likely the result of weather patterns. However, for the foreseeable future, point-source observations will be the only viable method to extract brown dwarf and exoplanet spectra. Models have been able to reproduce the observed variability, but ground truth observations are required to verify their results. To that end, we provide visual and near-infrared spectra of Jupiter and Saturn obtained from the \emph{Cassini} VIMS instrument. We disk-integrate the VIMS spectral cubes to simulate the spectra of Jupiter and Saturn as if they were directly imaged exoplanets or brown dwarfs. We present six empirical disk-integrated spectra for both Jupiter and Saturn with phase coverage of $1.7^\circ$ to $133.5^\circ$ and $39.6^\circ$ to $110.2^\circ$, respectively. To understand the constituents of these disk-integrated spectra, we also provide end member (single feature) spectra for permutations of illumination and cloud density, as well as for Saturn's rings. In tandem, these disk-integrated and end member spectra provide the ground truth needed to analyze point source spectra from extrasolar gas giants and brown dwarfs. Lastly, we discuss the impact that icy rings, such as Saturn's, have on disk-integrated spectra and consider the feasibility of inferring the presence of rings from direct imaging spectra. 

\end{abstract}



\section{Introduction}
\label{sec:intro}

\subsection{Brown Dwarfs and Gas Giants}
\label{subsec:bdagg}
Brown dwarfs are substellar objects that are not massive enough to sustain the fusion of hydrogen into helium but are massive enough to sustain deuterium fusion ($13-75$ $\mathrm{M_J}$). Their temperatures range from over $2000$ K for the hottest L-class brown dwarfs down to $\sim250$ K for the coldest Y-class brown dwarf discovered to date \citep{Luhman_2014}. Due to their relatively low temperatures, brown dwarfs have atmospheres that allow for the condensation of molecules into clouds. Hot L-class brown dwarfs ($1300-2200$ K) have atmospheric spectra characterized by strong water vapor opacity, carbon dioxide, metallic hydrides, alkali metals, and optically think clouds thought to be composed of silicates and iron \citep{Kirkpatrick_1999, Marley_2014}. For T-class brown dwarfs ($600-1300$ K), these signatures are dominated by prominent methane and water vapor absorption bands \citep{Kirkpatrick_1999, Burgasser_2002a}. One model for the L to T spectral class transition is marked by a shift from a largely uniform cloud layer to a mix of cloudy and cloudless regions, which eventually results in atmospheres potentially devoid of clouds \citep{Burgasser_2002b, Marley_2010}. However, this specific model appears inconsistent with recent observations \citep[e,g,][]{Apai_2013, Lew_2020}, although by some mechanism the T dwarfs are indeed mostly devoid of clouds. The spectra of cool Y-class brown dwarfs (less than $600$ K) show even more pronounced methane and water absorption bands and the coolest members even show evidence of ammonia \citep{Cushing_2011}. Extrasolar gas giant planets cover this same temperature range, and are theorized to show similar molecular and cloud signatures \citep{Sudarsky_2000, Sudarsky_2003, Fortney_2020}.

Understanding how clouds form and evolve is a primary objective of brown dwarf and extrasolar gas giant research (see Question B3 in the white paper by \citet{Apai_2017}).  Clouds are of first-order importance in these atmospheres, as they dramatically impact radiation transport and the atmospheric depth that one probes in spectral observations \citep{Marley_1999}. The energy radiated from a brown dwarf's atmosphere taps into the energy reservoir from past and present gravitational contraction. The emitted thermal radiation is mediated by the absorption due to molecules and cloud opacity in brown dwarf atmospheres.  Clouds act as a barrier to this radiation, with patchy clouds yielding non-uniform emission across the disk. Over the past several years, time-domain studies of the varied thermal emission of brown dwarfs have expanded the matured \citep[e.g.][]{Metchev_2015, Zhou_2016, Yang_2016, Cushing_2016, Apai_2017}, solidifying the view brown dwarfs have non-homogeneous photospheres \citep{Tan_2019, Tan_2021}.

It is likely that all brown dwarfs are variable to some degree. The most variable appear to be those at the L-to-T transition, where cloud opacity is changing dramatically, suggesting a close connection between spatially heterogeneous cloud opacity and variability. A recent review of time-domain observations is provided by \citet{Biller_2017}, and impressive overview of models and observations is found in \citet{Zhang_2020}. In a manner analogous to the brown dwarfs, Jupiter and Saturn also radiate thermal energy due to leftover heat from their formation \citep{Aumann_1969}. These planets also have strongly non-uniform surfaces, marked by regions of thick and thin clouds. These atmospheres have been studied in increasing detail over the past several decades, in particular via space missions. The Cassini Mission flyby of Jupiter in 2000 and orbit around Saturn during 2004-2017 provided fresh data from new viewing geometries that allow us a unique probe of the reflected and thermal flux from these non-homoegenous atmospheres. In this paper we study these solar system spectra in the context of exoplanets and brown dwarfs to draw connections between these close-by and more distant objects.

\subsection{Direct Imaging Spectroscopy}
\label{subsec:direct}
Progress in exoplanet atmosphere characterization via high-contrast imaging has been challenging due to technological limitations, but refined techniques and the next era of space telescopes show promise of significant progress in the field. Recent reviews on direct imaging include \citet{Traub_2010} and \citet{Biller_2018}. There is a strong interest in the direct imaging and subsequent characterization of all varieties of exoplanets. While brown dwarfs can typically be observed in isolation, without the bright glow of a parent star, this is not true for exoplanets. The James Webb Space Telescope (JWST) and Nancy Grace Roman Space Telescope (formerly WFIRST) will be fitted with coronagraphs, which will enable direct imaging of exoplanets near bright stars. In addition, the Astro2020 decadal survey recommended a large $\sim$6-meter class space telescope with the driving science case to image and characterize exoplanets Earth-exoplanets with habitable zones of nearby Sunlike stars \citep{Decadal_2020}.  With the influx of direct imaging data expected in the coming decades, it is imperative that the exoplanet and brown dwarf communities are able to interpret this data using tools and modeling frameworks trained with ground truth observations from the solar system.  

Trying to assess the reflection and emission from solar system planets in the exoplanet context has expanded as well.  Recently, spectral analysis studies have been conducted for Earth \citep{Livengood_2011, Jiang_2018, Gu_2021} and Neptune \citep{Simon_2016}, providing solar system analogs for terrestrial planets and ice giants, respectively. Similar studies have been done for Jupiter \citep{Gelino_2000, Ge_2019}, which along with Saturn, are our only spatially resolvable proxies for brown dwarfs and extrasolar gas giants. These studies present photometric variability of an unresolved Jupiter, with the most significant variability in the infrared, the result of patchy clouds. Our study builds upon these results in four key ways. First, we include spectra of Saturn in addition to Jupiter, which provides an additional analog to extrasolar gas giants and brown dwarfs. Second, previous studies have very low spectral resolution, two bands in \citet{Gelino_2000} and twelve bands in \citet{Ge_2019}, so high resolution spectral analogs are badly needed. We provide disk-integrated spectra of Jupiter and Saturn with 330 spectral channels across the visible and near-infrared. Third, we present end member spectra of Jupiter and Saturn (see Section \ref{subsec:endmember}) in addition to disk-integrated spectra. And finally, we make our spectra publicly available for use in training atmosphere models.

\begin{deluxetable*}{lllrrr}[t]
\centering
\label{tab:observations}
\tablecaption{VIMS spectral cubes selected for this study. All data presented here were retrieved from the NASA Planetary Data System (PDS) archive, except the resolution values marked with an asterisk, which were not available in PDS and we calculated ourselves. Color images created from each cube can be found in Figure \ref{fig:phases}.}
\tabletypesize{\scriptsize}
\tabletypesize{\small}
\tablewidth{0pt}
\tablehead{
\colhead{Target} & \colhead{Cube Name} & \colhead{Acquisition Date} & \colhead{VIS, IR Exposure Duration (s)} &  \colhead{Resolution (km/pixel)} & \colhead{Phase Angle ($^\circ)$}}
\decimals
\startdata
Jupiter & V1355313318\_3 & 2000 Dec 12 & 640, 20 & 9660 & 1.66 \\
 & V1353677179\_2 & 2000 Nov 23 & 640, 20 & 17,478* & 15.19 \\
 & V1357119162\_1 & 2001 Jan 2 & 640, 20 & 5011 & 76.96 \\
 & V1357335218\_1 & 2001 Jan 4 & 640, 20 & 5455 & 89.26 \\
 & V1357767306\_1 & 2001 Jan 9 & 640, 20 & 6872 & 107.27 \\
 & V1359577908\_1 & 2001 Jan 30 & 640, 20 & 15,536* & 133.52 \\
\hline
Saturn & V1564773481\_1 & 2007 Aug 2 & 1280, 20 & 1862 & 39.64 \\
 & V1565414411\_1 & 2007 Aug 10 & 1280, 20 & 2031 & 49.61 \\
 & V1565931915\_1 & 2007 Aug 16 & 1280, 20 & 1883 & 57.76 \\
 & V1469259344\_1 & 2004 Jul 23 & 1280, 120 & 3304 & 91.90 \\
 & V1518276136\_1 & 2006 Feb 10 & 2560, 40 & 1925 & 101.07 \\
 & V1517685790\_1 & 2006 Feb 3 & 2560, 40 & 1990 & 110.20 \\
\hline
\enddata
\end{deluxetable*}

In this paper we present visual and near infrared disk-integrated spectra of Jupiter and Saturn at various phase angles, defined hereafter as the sun-planet-\emph{Cassini} angle. In Section \ref{sec:methods} we discuss the process of selecting and reducing the spectral cubes used in this study. In Sections \ref{sec:jupiter} and \ref{sec:saturn} we present the disk-integrated and end member spectra for Jupiter and Saturn, respectively. The disk-integrated spectra are proxies for point-source spectra of extrasolar gas giants obtained via direct imaging. The end member spectra for different combinations of illumination and cloudiness will be useful when attempting to deconstruct point-source spectra into constituent atmospheric features. In Section \ref{sec:saturn}, we also discuss the ability to discern the presence of icy rings from point source spectra. Lastly, in Section \ref{sec:conclusions}, we conclude by reiterating how this study improves upon previous work in the field and comment on potential future work in exo-ring detection using the results presented here.

\begin{figure*}[!h]
\begin{center}
\includegraphics[width=15.4cm]{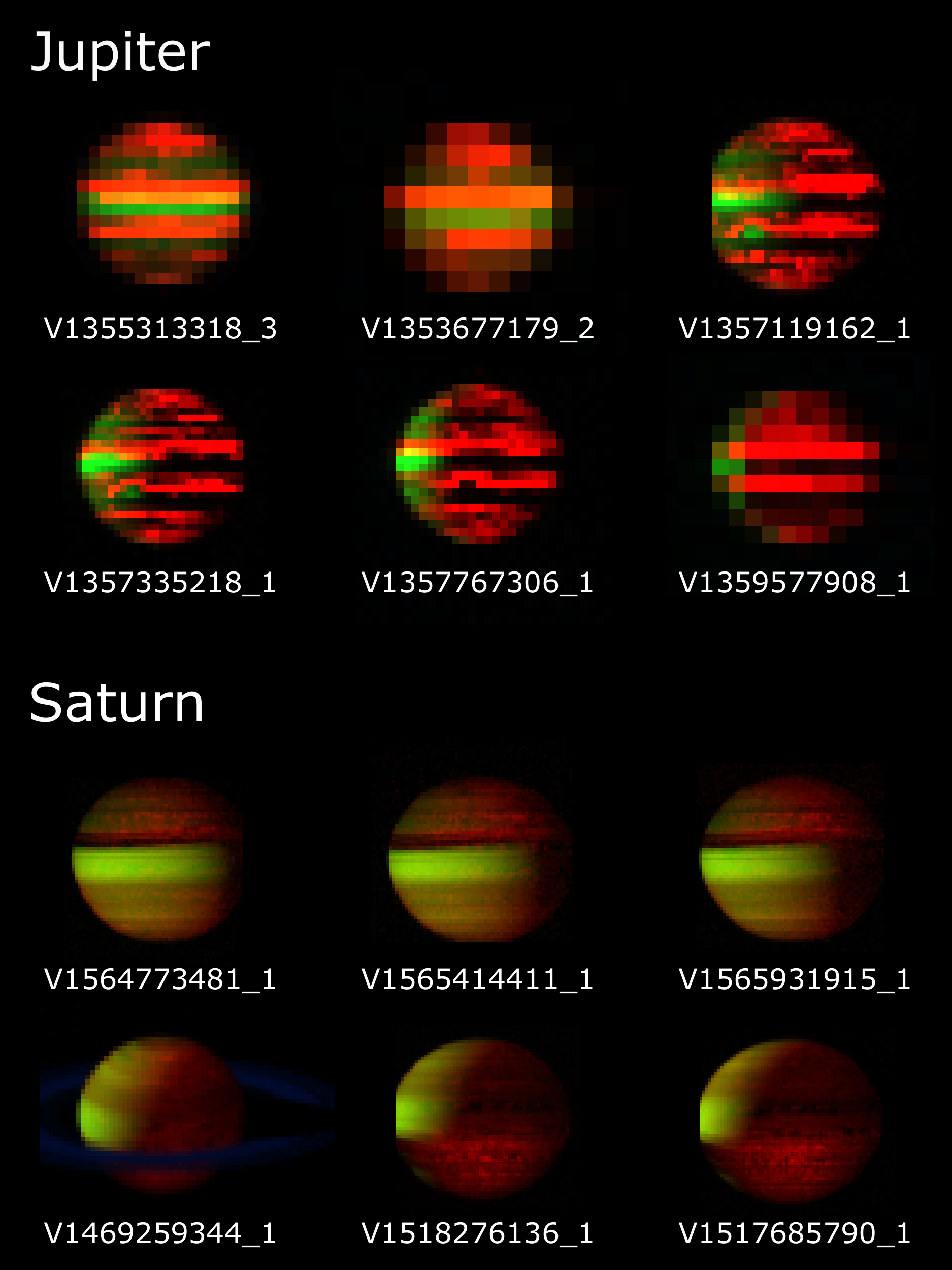}
\end{center}
\caption{Color images of all twelve cubes presented in this study with green mapped to 2.0 $\mu m$, red mapped to 5.0 $\mu m$, and blue mapped to 2.4 $\mu m$. In this color scheme, green represents reflected light, red represents thermal emission, and blue indicates the presence of water. The images of Saturn were created from the cubes before the ring removal algorithm was implemented. Note the lack of blue in all images except for V1469259344\_1, in which Saturn's rings, primarily composed of water ice, are clearly visible in blue.}
\label{fig:phases}
\end{figure*}

\section{Observations \& Methods}
\label{sec:methods}

\subsection{Cassini VIMS Observations}
\label{subsec:vims}

Spectral data comes from the \emph{Cassini} Visual and Infrared Mapping Spectrometer (VIMS) instrument \citep{Brown_2004}, which acquired individual point spectra with 352 spectral channels between 0.35 and 5.2 $\mu m$. VIMS is composed of two separate optical and detector assemblies -- the visual half, or VIMS-V, and the infrared half VIMS-IR -- that employ a common pointing and spacecraft interface such that they operate as if they were a single instrument. Using a targeting mirror, VIMS was able to build spectral image cubes up to 64x64 pixels in size, providing adequate spatial resolution for this study. \emph{Cassini} VIMS obtained over 14,000 cubes during its flyby of Jupiter and over 550,000 cubes of Saturn's disk during its 13 years in orbit. We compiled cubes suitable for this study based on three criteria:
\begin{enumerate}[label=(\roman*)]
\item The image showed the entire disk of the target planet in order to obtain complete disk-integrated spectra
\item There were no pixels on the target planet's disk that were saturated due to overexposure
\item There were no objects or artifacts that would affect the spectra, such as a transiting moon or severe cosmic ray spikes
\end{enumerate}

Under these constraints, the number of suitable cubes for both Jupiter and Saturn dropped to only a few dozen each, owing to the frequent overexposure of some wavelengths and the nature of \emph{Cassini}'s orbit, which rarely allowed for full disk imaging of Saturn. These remaining cubes were clustered in groups of images taken in quick succession at similar phase angles, with many phase angles absent. Thus, full phase angle coverage is not possible without relaxing the selection criteria. In the end, six cubes each were selected for Jupiter and Saturn so that they provided the most complete phase angle coverage possible for each planet and for the two combined. Color images of the cubes selected for this study are shown in Figure \ref{fig:phases} and important observational data can be found in Table \ref{tab:observations}.

\subsection{Data Reduction}
\label{subsec:reduction}

We retrieved raw VIMS cubes from the NASA Planetary Data System (PDS) archive and reduced them using the pipeline described in \citet{2013ApJ...777..161B}. The processing proceeds as follows:
\begin{enumerate}[]
\item On-board the spacecraft, for each line VIMS observes it measures a background level. Raw cubes as downloaded from the spacecraft have the average value of this background measurement already subtracted from them. However since the on-board backgrounds have proven to be noisy, we first add the mean background back in to each pixel for the visual half of VIMS and the full measured background back in for the infrared half. Then for each cube we create a linear fit to the instrument-measured background observations acquired at the end of each row in the cube as a function of time and subtract the linearly interpolated background from each pixel. This method has proven to be much more robust against cosmic ray spikes and the slowly varying background level as compared to the on-board subtraction.
\item We begin the automatic pipeline process by marking those pixels that have saturated.
\item Next, the pipeline employs a despiking algorithm to identify pixels whose flux is dominated by cosmic ray hits. We assign the offending pixels an interpolated value based
on a 3-D polynomial fit of the twenty-six nearest neighbor pixels. Results of the automated despiker compare favorably to manually despiked cubes for extended objects like Jupiter and Saturn, although the despike algorithm does incorrectly flag specular reflections off of Titan lakes as cosmic ray hits \citep{2013ApJ...777..161B}. 
\item We divide through by a spatial flatfield that was taken on the ground before launch at each wavelength. Lacking an instrumental mechanism for generating an in-flight flatfield, we tried using hundreds of uniformly illuminated science observation cubes to generate an empirical flatfield, but found the result inferior to the ground-based calibration.
\item We convert to specific energy ($I_\lambda$) using the instrument's spectro-radiometric response function $R(\lambda)$ in photons per analog-to-digital data number ($DN$) \citep{Brown_2004}:
\begin{equation}
    I_\lambda = \frac{DN}{\tau}\times R(\lambda)\times \frac{hc}{\lambda}\times \frac{1}{A\Omega \delta \lambda}
\end{equation}
where $\tau$ is the exposure time in seconds, $h$ is Planck's constant, $c$ is the speed of light, $\lambda$ is the wavelength of this particular VIMS channel, $\delta \lambda$ is the size of the wavelength bin for this VIMS channel, $A$ is the area of the VIMS mirror, and $\Omega$ is the solid angle subtended by a pixel in steradians.
\item Using SPICE \citep{Acton_1999} data provided by the NASA Jet Propulsion Laboratory (JPL), we calculate the latitude, longitude, x- and y-direction resolution, phase angle, incidence angle, emission angle, and north azimuth of every point individually. These data are stored in the cube backplane. Since VIMS is a spot-scanner \citep{Brown_2004}, we are able to compensate for spacecraft pointing changes during an exposure. The SPICE geometry kernel files also contain knowledge of the spacecraft position relative to the target planet, which we ultimately use to convert observed specific energy into the net spectral flux of the planet.
\end{enumerate}

Some aspects of the VIMS team data reduction pipeline were hardcoded for reducing cubes obtained during \emph{Cassini}'s time orbiting Saturn. Therefore, for the background subtraction in the visible wavelengths we used a separate but equivalent procedure when reducing Jupiter observations. To calibrate for visible wavelengths, we calculated the median value of the background (non-disk) pixels for each spectral band and then subtracted that value from all pixels in that band.

\subsection{Disk-Integrated Spectra}
\label{subsec:integration}

Disk integration of Jupiter's spectral cubes was straightforward. For both spectral radiance and apparent reflectance (I/F), we simply coadded all pixels on the disk and divided by the number of pixels coadded. However, Saturn was more challenging, since information on the geometry of the rings is not present within the cube backplanes. Therefore, we manually masked out the pixels containing the rings and their shadows from the image. We replaced each missing pixel by swapping it with the pixel with the nearest incidence angle value. This process was done separately for VIMS-V and VIMS-IR since they had slightly different viewing geometries. Once the rings were synthetically removed from the images, we disk-integrated them as we did for Jupiter.  Integrated images of Saturn with rings included were also produced and the comparison to the ``ringless" Saturn is discussed in Section \ref{sec:saturn}. This process provides spatially resolved examples of a planet with and without rings since some planets and brown dwarfs will host impressive ring systems while others will not.

\begin{figure*}[!tb]
\begin{center}
\includegraphics[width=12.5cm]{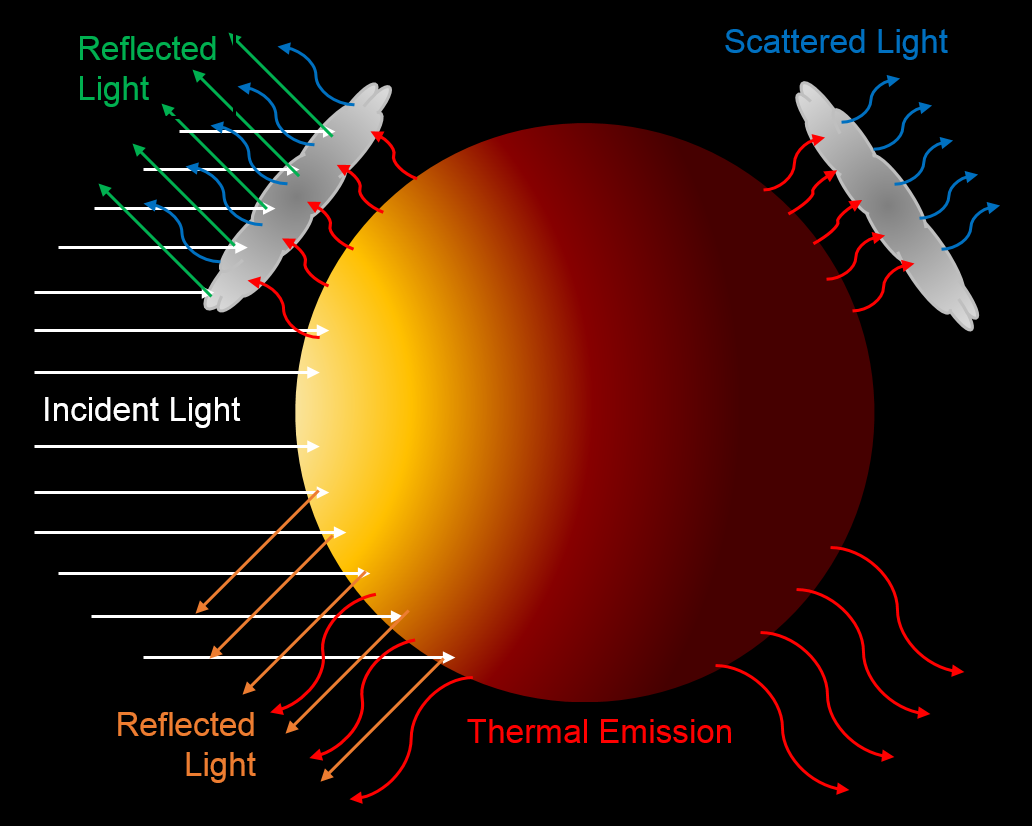}
\end{center}
\caption{Visual representation of the four end members outlined in Section \ref{subsec:endmember}. Incident solar light (white) can either be reflected by dense clouds (green) or cloudless regions (orange). Thermal emission (red) are either directly emitted or scattered by cloud layers (blue).}
\label{fig:cartoon}
\end{figure*}

\subsection{End Member Spectra}
\label{subsec:endmember}

We obtained end member, or single feature, spectra by selecting an individual pixel of the target feature and determining its radiance and apparent reflectance (I/F) values across all VIMS spectral channels, once again accounting for the difference in viewing geometry between VIMS-V and VIMS-IR. For each end member, we chose three such pixels and averaged them to reduce any noise present in a single pixel. Four end members are common to Jupiter and Saturn:
\begin{enumerate}[]
\item \textbf{Dayside and Cloudy}: Regions that are on the illuminated area of the disk and have thick clouds that both reflect incident light and block thermal emission from the planet (upper left in Figure \ref{fig:cartoon}). On Jupiter, these are the cloud-dense zones on the illuminated side of the disk (bright green in Figure \ref{fig:phases}). On Saturn, this is the illuminated part of the cloudy region straddling its equator (yellow in Figure \ref{fig:phases}).
\item \textbf{Dayside and Clear}: Regions that are on the illuminated area of the disk and have few clouds to block thermal emission (lower left in Figure \ref{fig:cartoon}). On Jupiter, these are the cloud-free belts on the illuminated side of the disk (red on the illuminated side in Figure \ref{fig:phases}). On Saturn, this is the darker region just above the bright equatorial zone (reddish-brown in Figure \ref{fig:phases}).
\item \textbf{Nightside and Cloudy}: Regions that are on the dark side of the disk and have thick clouds that block thermal emission from the planet (upper right in Figure \ref{fig:cartoon}). On Jupiter, these are the cloud-dense zones on the dark side of the disk (black in Figure \ref{fig:phases}). On Saturn, this is the dark part of the cloudy region straddling its equator (red in Figure \ref{fig:phases}).
\item \textbf{Nightside and Clear}: Regions that are on the dark side of the disk and have fewer clouds to block thermal emission (lower right in Figure \ref{fig:cartoon}). On Jupiter, these are the belts on the dark side of the disk (red on the illuminated side in Figure \ref{fig:phases}), which have thinner clouds than the zones. On Saturn, this is the darker region just above the equatorial zone (dark red in Figure \ref{fig:phases}).
\end{enumerate}

For Saturn, we produced two additional end member spectra for the A Ring and B Ring. All end member spectra from Jupiter and Saturn were obtained from cubes V1357335218\_1 and V1469259344\_1, respectively.

\begin{figure*}[!tb]
\begin{center}
\includegraphics[width=\textwidth]{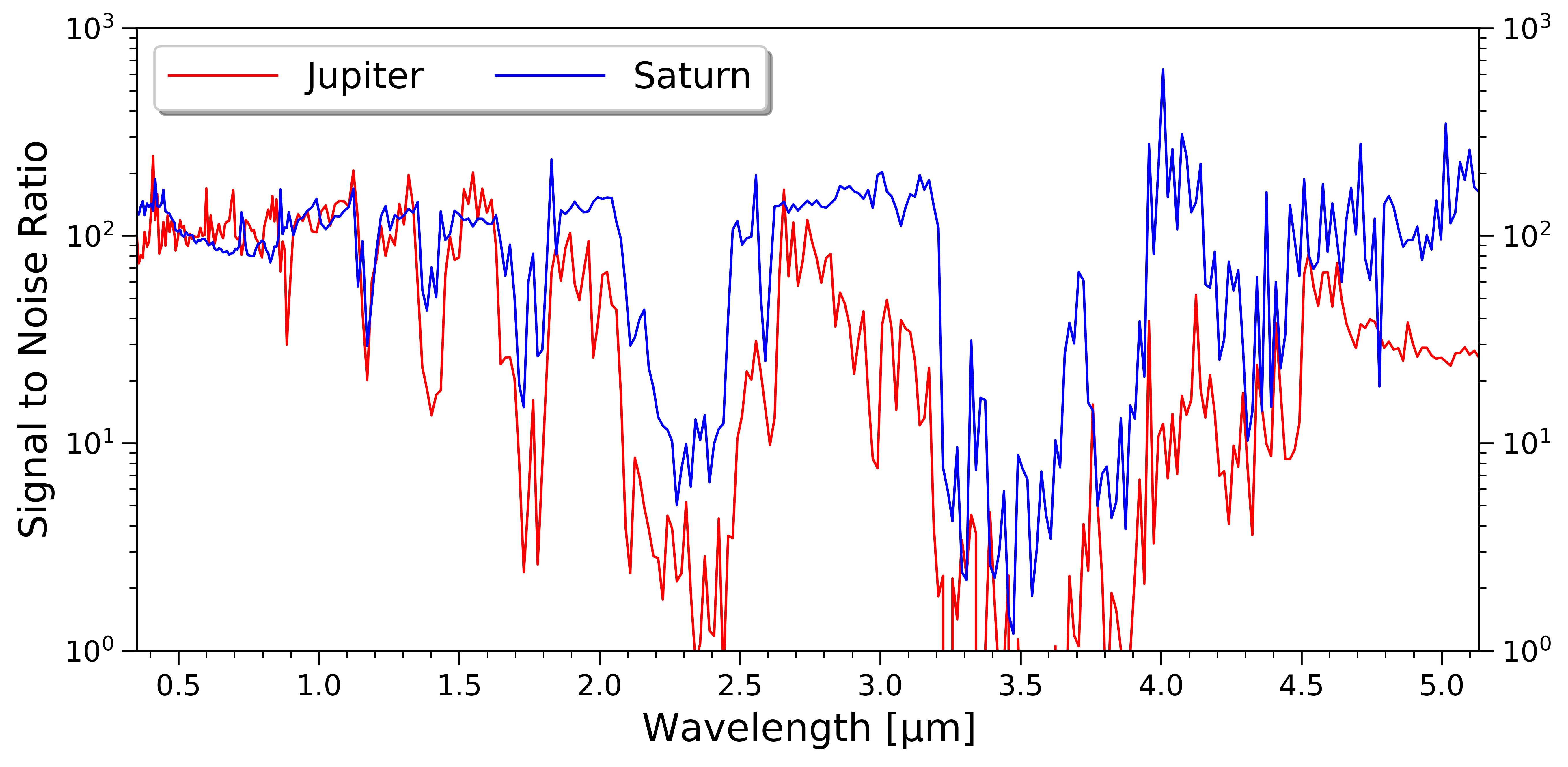}
\end{center}
\caption{Signal to noise ratio as a function of wavelength for both Jupiter and Saturn. As expected, the signal to noise ratio is very low in the major absorption bands. The data used to generate this figure can be found in the supplementary information.}
\label{fig:stn}
\end{figure*}

\subsection{Signal to Noise Ratio Calculation}
\label{subsec:stn}

Uncertainties in the spectra are dominated by astrophysical aspects (i.e. angular position, small atmospheric variations) rather than instrumental ones. Therefore, viewing Jupiter or Saturn at the same phase angle, but at different times (thus different angular positions) will result in slightly different spectra. To account for these variations, we averaged the spectra of four nearby cubes for Jupiter and five such cubes for Saturn. The phase variations in these sets are $0.43^\circ$ for Jupiter and $0.68^\circ$ for Saturn. We calculated the means and standard deviations for both sets and divided the mean spectra by the standard deviation spectra (taken here to be the noise) to get a signal to noise ratio for both planets. Our results are shown in Figure \ref{fig:stn}.

\section{Jupiter}
\label{sec:jupiter}

\subsection{Results}
\label{subsec:jupresults}

\begin{figure*}[!tb]
\begin{center}
\begin{subfigure}{\textwidth}
\includegraphics[width={\textwidth}]{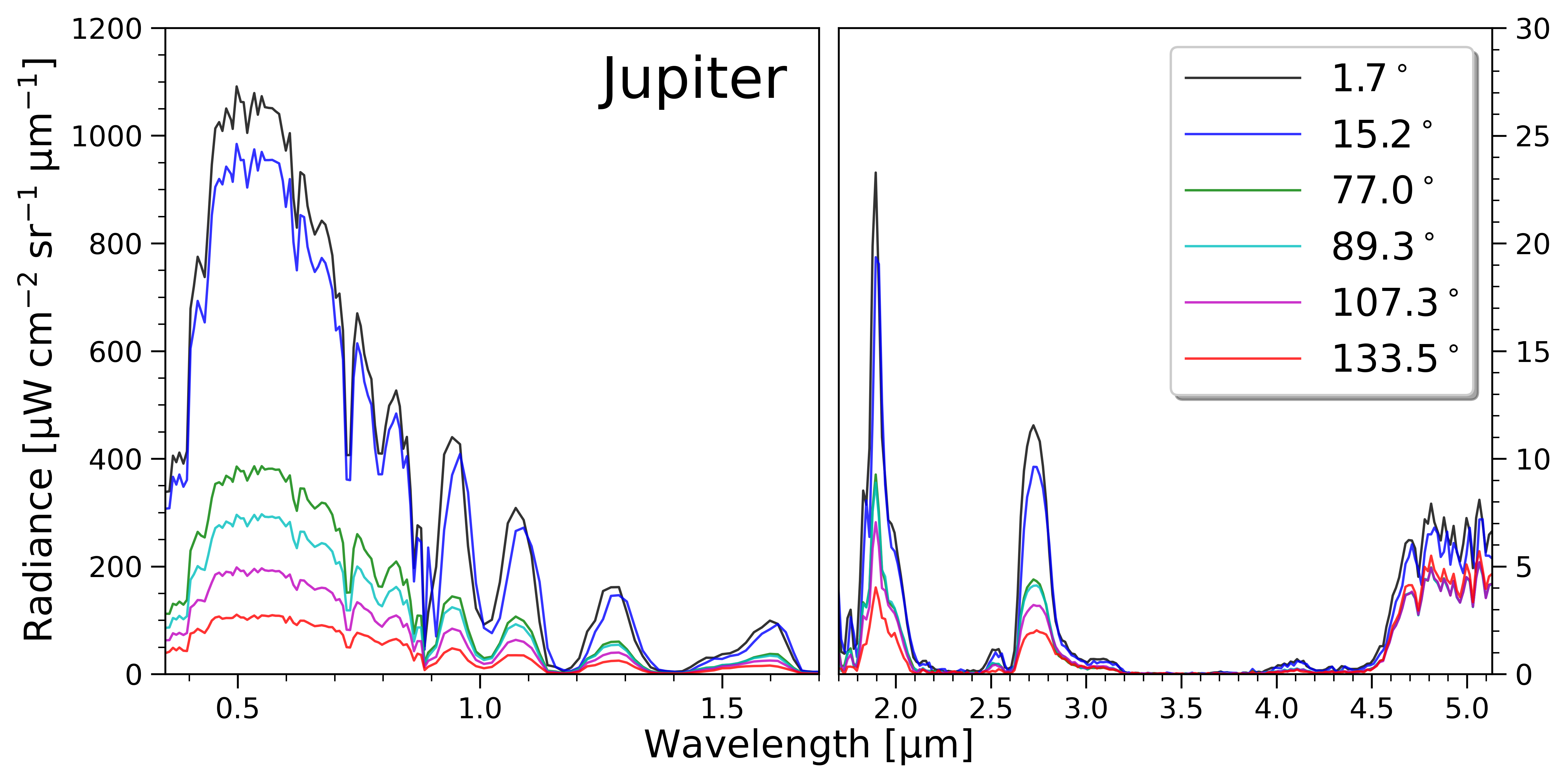}
\end{subfigure}
\begin{subfigure}{\textwidth}
\includegraphics[width={\textwidth}]{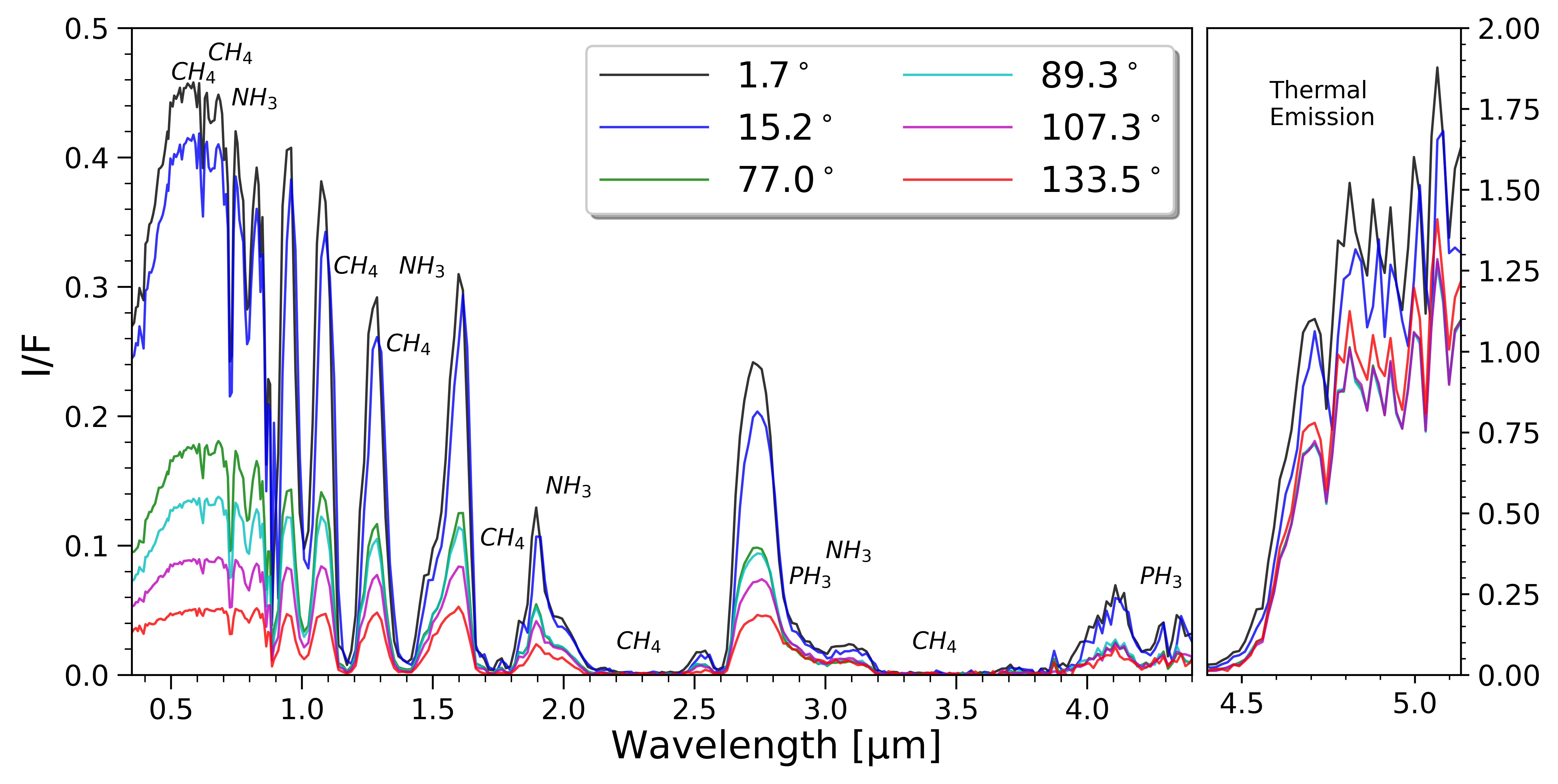}
\end{subfigure}
\end{center}
\caption{\emph{Top}: Disk-integrated spectra of Jupiter in units of flux for six phase angles ranging from $1.7^\circ$ to $133.5^\circ$. \emph{Bottom}: Same as above, but in units of I/F. Prominent absorption bands are labelled. Absorption bands in the visual and infrared wavelengths are taken from \citet{Vdovichenko_2021} and \citet{Baines_2005}, respectively. The source data for this figure can be found in the supplementary information.}
\label{fig:jupiter}
\end{figure*}

\begin{figure*}[!tb]
\begin{center}
\begin{subfigure}{\textwidth}
\includegraphics[width={\textwidth}]{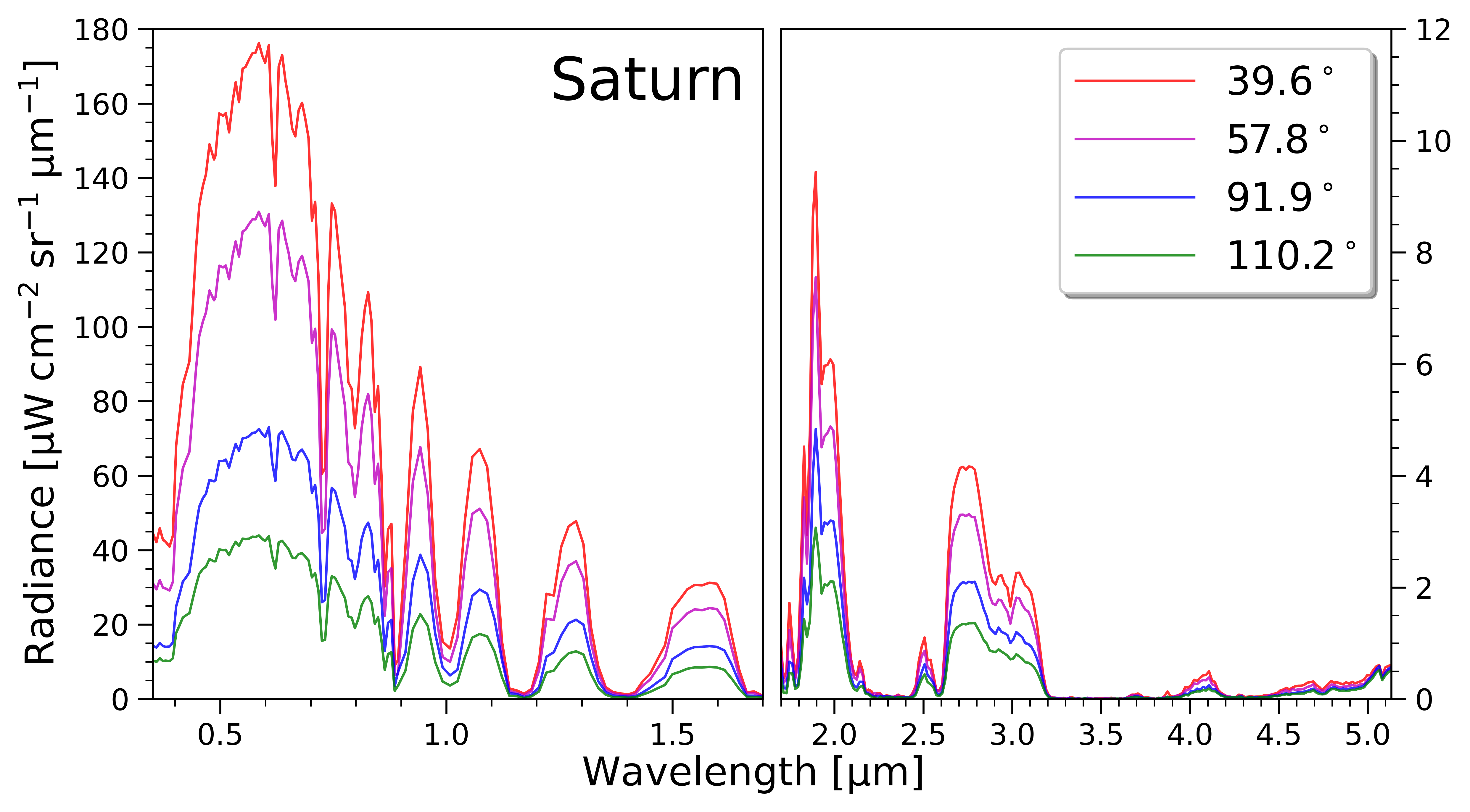}
\end{subfigure}
\begin{subfigure}{\textwidth}
\includegraphics[width={\textwidth}]{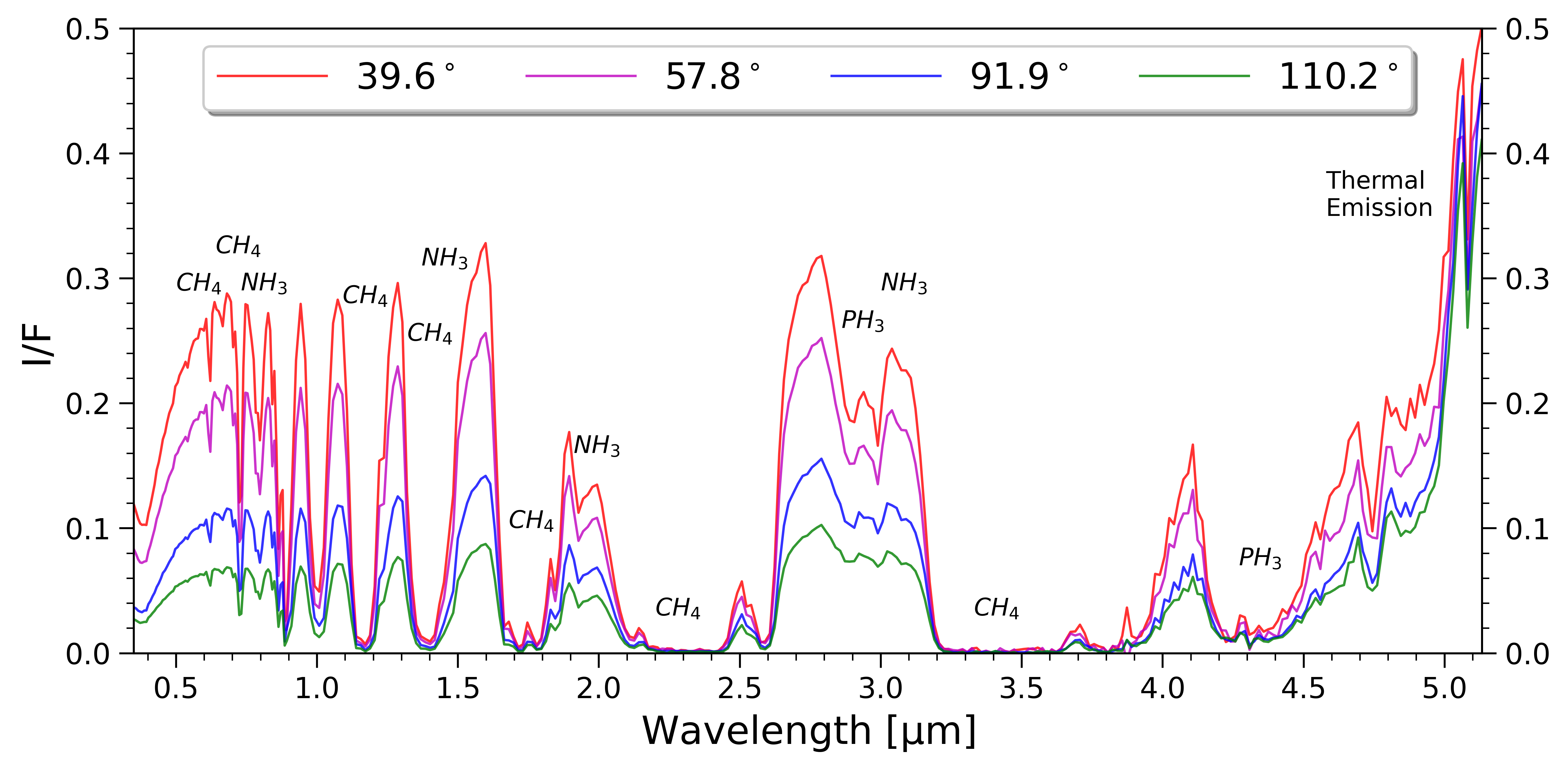}
\end{subfigure}
\end{center}
\caption{\emph{Top}: Disk-integrated spectra of Saturn in units of flux for four phase angles ranging from $39.6^\circ$ to $110.2^\circ$. Due to the nature of Cassini's orbit around Saturn, full disk, full phase spectra were impossible to obtain. \emph{Bottom}: Same as above, but in units of I/F. Once again, prominent absorption bands, courtesy of \citet{Vdovichenko_2021} and \citet{Baines_2005}, are labelled. The data used to generate this figure, as well as for two additional phase angles, can be found in the supplementary information.}
\label{fig:saturn}
\end{figure*}

Disk-integrated spectra for six phase angles of Jupiter are shown in Figure \ref{fig:jupiter}. The visual and near infrared (below 3.5 $\mu m$) spectrum of Jupiter is dominated by reflected solar light with absorption bands primarily from methane ($CH_{4}$), phosphine ($PH_{3}$), and ammonia ($NH_{3}$). Of note is the ammonia absorption band at 3.0 $\mu m$, which is much more pronounced in Jupiter, leading to a single peak structure in the spectrum as opposed to the dual peak structure seen in Saturn (see Figure \ref{fig:saturn}). Beyond 3.5 $\mu m$ the influence of solar light wanes dramatically and the spectrum is now primarily thermal emission. Therefore, the phase-dependent variations are no longer present in this regime. Despite the thermal radiation being minimal in comparison to reflected light in flux, it dominates the I/F spectrum (note the axis changes in both spectra of Figure \ref{fig:jupiter}).

\begin{figure*}[!tb]
\begin{center}
\includegraphics[width={\textwidth}]{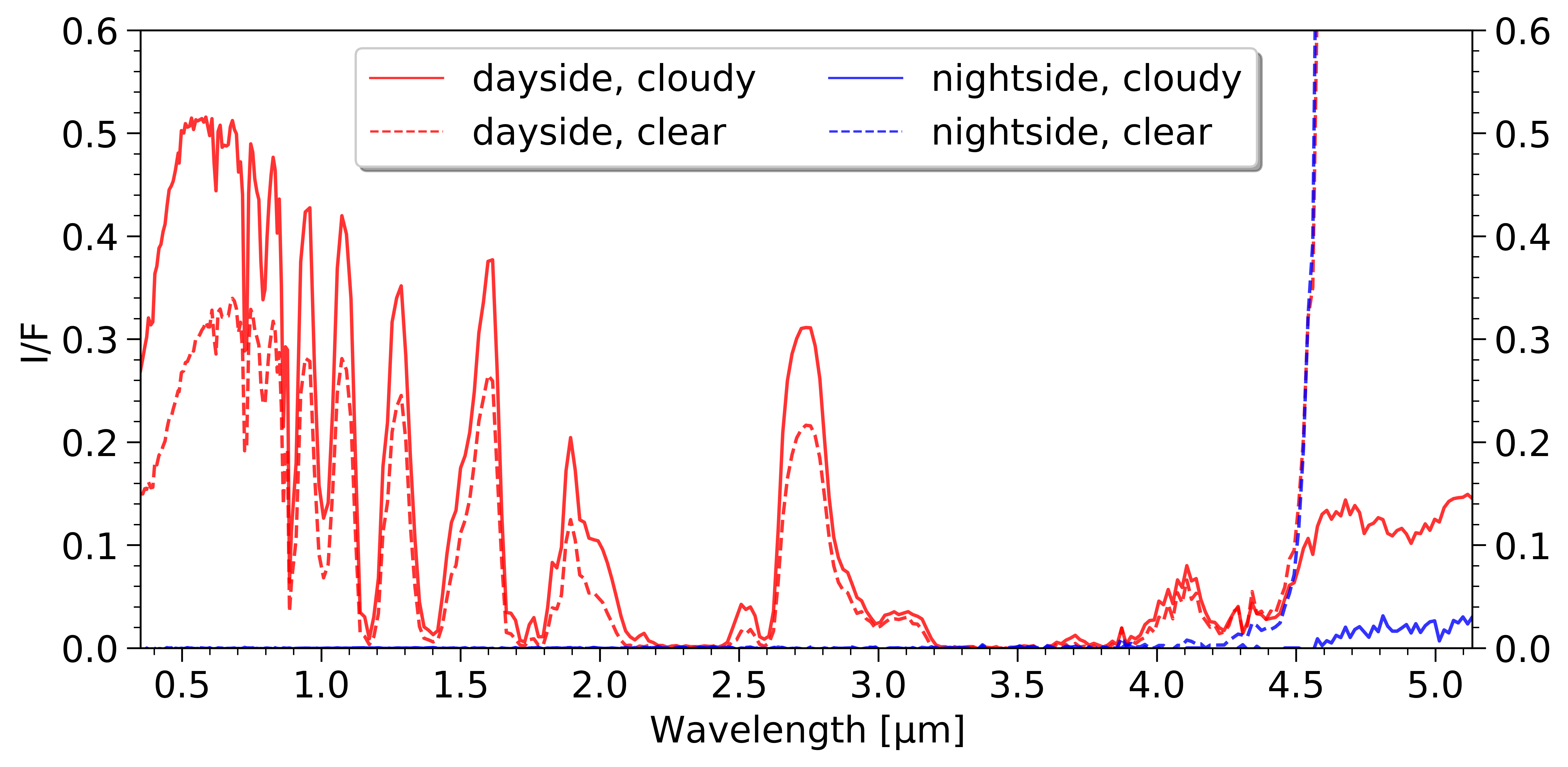}
\end{center}
\caption{Jupiter end member spectra obtained from cube V1357335218\_1 (see Figure \ref{fig:phases}). The data used to generate this figure can be found in the supplementary information.}
\label{fig:jupiterem}
\end{figure*}

Jupiter end member I/F spectra are shown in Figure \ref{fig:jupiterem}. In the visual and near-infrared, both the dayside and nightside spectra are structurally similar, however their cloudy and clear members diverge dramatically in the mid-infrared. This dichotomy is the result of the extreme banding present in Jupiter's atmosphere, which can be seen visually in Figure \ref{fig:phases}. Jupiter's zones are particularly cloud-dense while the clouds are relatively sparse in their counterpart, the belts. The result is a significant percentage of thermal emission in zones are blocked, while thermal energy can more easily escape from belts. Beyond 4.5 $\mu m$, thermal emission dominates, so there should be no difference between the clear spectra (dashed lines in Figure \ref{fig:jupiterem}) in this wavelength regime.

\subsection{Discussion}
\label{subsec:jupdiscussion}

The intensity variations observed in Figure \ref{fig:jupiter} are the combination of phase-dependent illumination with smaller effects from Jupiter's scattering phase function. As the primary purpose of this study is to provide the exoplanet and brown dwarf communities with disk-integrated spectra as proxies for direct imaging observations, the data was not altered to account for any of these effects. A full analysis of the effects of phase-dependent scattering in Jupiter's atmosphere is beyond the scope of this study. However, studies by \citet{Mayorga_2016} and \citet{Heng_2021} provide a robust analysis of phase-dependent effects in Jupiter's atmosphere, specifically in the context of comparison to direct imaging data.

Over the past several years, atmosphere models that aim to assess the temperature structure, composition, and reflection/emission spectra of directly imaged planets, as well as ``retrieval'' codes that aim to turn reflection spectra into atmospheric constraints, have been developed in anticipation of upcoming spectra \citep{Lupu_2016, Nayak_2017, Macdonald_2018, Damiano_2020a, Damiano_2020b}. The disk-integrated spectra presented here will be vitally important for training such models with ground truth spectra from spatially resolvable solar system gas giants. Direct imaging requires that the target planet be further from the host star than other detection methods. Therefore, direct imaging spectroscopy is biased towards planets with modest incident fluxes, perhaps $\sim$1-5 AU, making the cool gas giants in our solar system ideal analogs. Planets as cool as Jupiter $T_{\rm eff}=125$ K are dominated by gaseous methane and ammonia clouds \citep{Sudarsky_2000}, so abundances of methane and other molecules can be determined by the depth of their absorption bands in reflection spectra. Therefore, future observing campaigns should have wavelength coverage and resolution to observe these absorption bands (e.g. the large methane bands centered on 2.3 $\mu m$ and 3.5 $\mu m$ and the ammonia band at 3.1 $\mu m$).

Since it emits thermal radiation, Jupiter also acts as a brown dwarf analog when viewed in the mid-infrared. Specifically, Jupiter is compositionally similar to ultra-cool Y-class brown dwarfs, so its utility as a brown dwarf analog increases as we probe cooler and cooler brown dwarfs such as WISE J035000.32-565830.2 \citep{Leggett_2016}, WISE J085510.83-071442.5 \citep{Luhman_2014, Skemer_2016, Morley_2018}, and others \citep{Leggett_2015}. As shown in Figure \ref{fig:jupiter}, this thermal emission is most prominent beyond 4.5 $\mu m$, where cool brown dwarfs are also known to be quite bright. The intensity of the thermal emission is highly dependent upon cloud density, which is apparent in Figure \ref{fig:jupiterem}. Jupiter's emission near 5 $\mu$m emerges almost exclusively from ``holes'' in the clouds, where cloud opacity is low. Y-class brown dwarf atmospheres may exhibit the same banded structure as Jupiter \citep{Zhang_2014}, making it an excellent test case for future brown dwarf observations. In addition, the end member spectra presented here can aid in constraining the relative amounts of cloudy and cloudless regions in their atmospheres.

\section{Saturn}
\label{sec:saturn}

\subsection{Results}
\label{subsec:satresults}

\begin{figure*}[!tb]
\begin{center}
\includegraphics[width={\textwidth}]{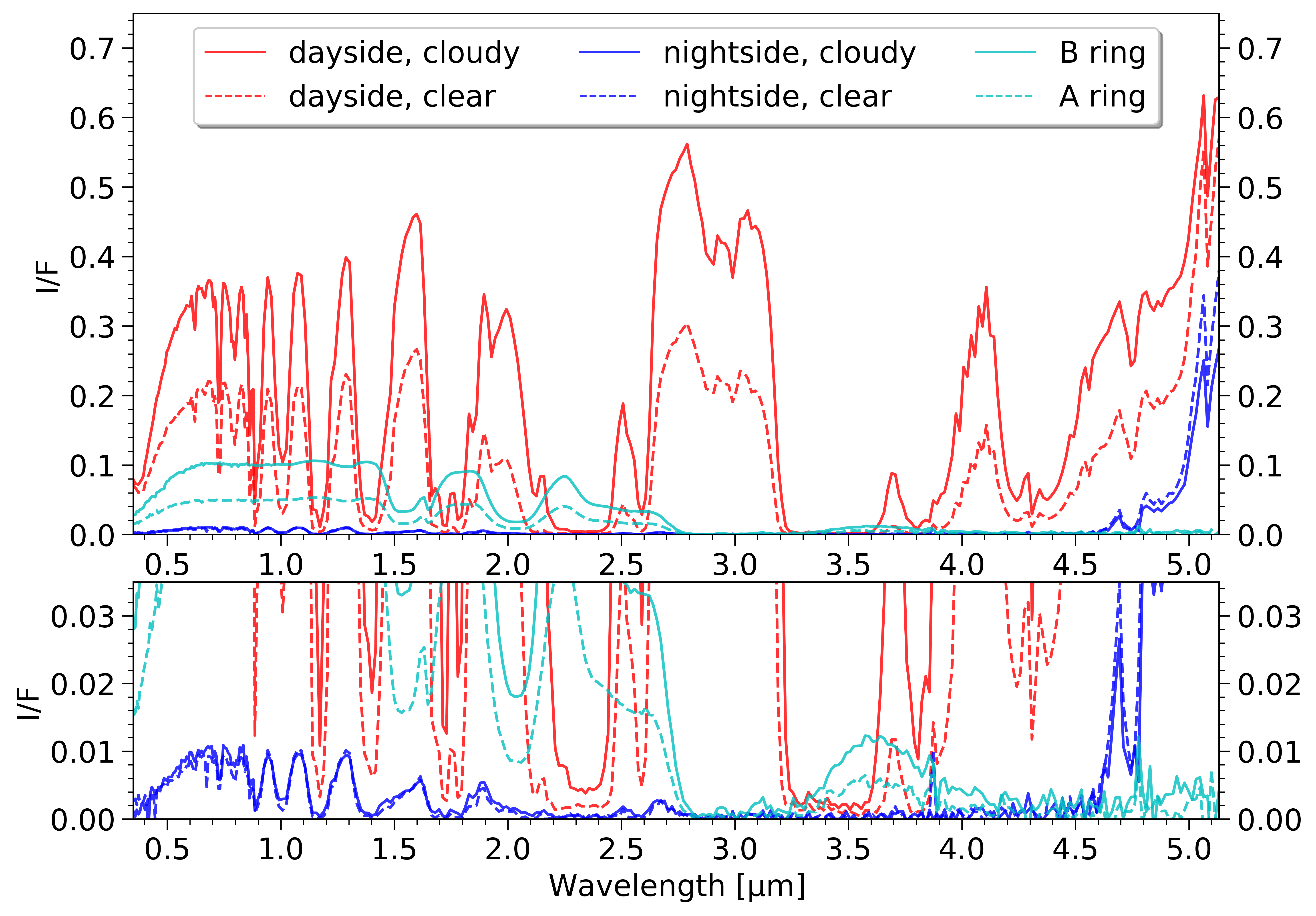}
\end{center}
\caption{\emph{Top}: Saturn end member spectra obtained from cube V1469259344\_1 (see Figure \ref{fig:phases}). \emph{Bottom}: Same as above but the y-scale has been adjusted to show detail. The features in the nightside spectra in the visible wavelengths are evidence of ringshine from Saturn's rings. The data used to generate this figure can be found in the supplementary information.}
\label{fig:saturnem}
\end{figure*}

Disk-integrated spectra for four phase angles of Saturn are shown in Figure \ref{fig:saturn}. As with Jupiter, the observed intensity variations are the combination of phase-dependent illumination and scattering phase function. Studies by \citet{Tomasko_1980} and \citet{Perez_2016} provide analysis of Saturn's atmospheric phase function. Saturn's visual and near-infrared spectrum is dominated by reflected light and many of the same methane, phosphine, and ammonia absorption bands as Jupiter, while the mid-infrared is dominated by thermal radiation. Owing to its much hazier atmosphere and cooler temperature, we do not see the same dramatic spike beyond 4.5 $\mu m$ that we do in Jupiter's spectrum. This difference could potentially indicate the severity of banding due to haziness of brown dwarfs and gas giants from point-source spectra alone. Saturn end member spectra are shown in Figure \ref{fig:saturnem}. Below 4.5 $\mu m$, the end members resemble those of Jupiter, but beyond 4.5 $\mu m$, we do not see the same wild divergence between cloudy and clear spectra. This lack of divergence in the infrared is once again the result of Saturn's hazy atmosphere. 

\begin{figure*}[!tb]
\begin{center}
\includegraphics[width={\textwidth}]{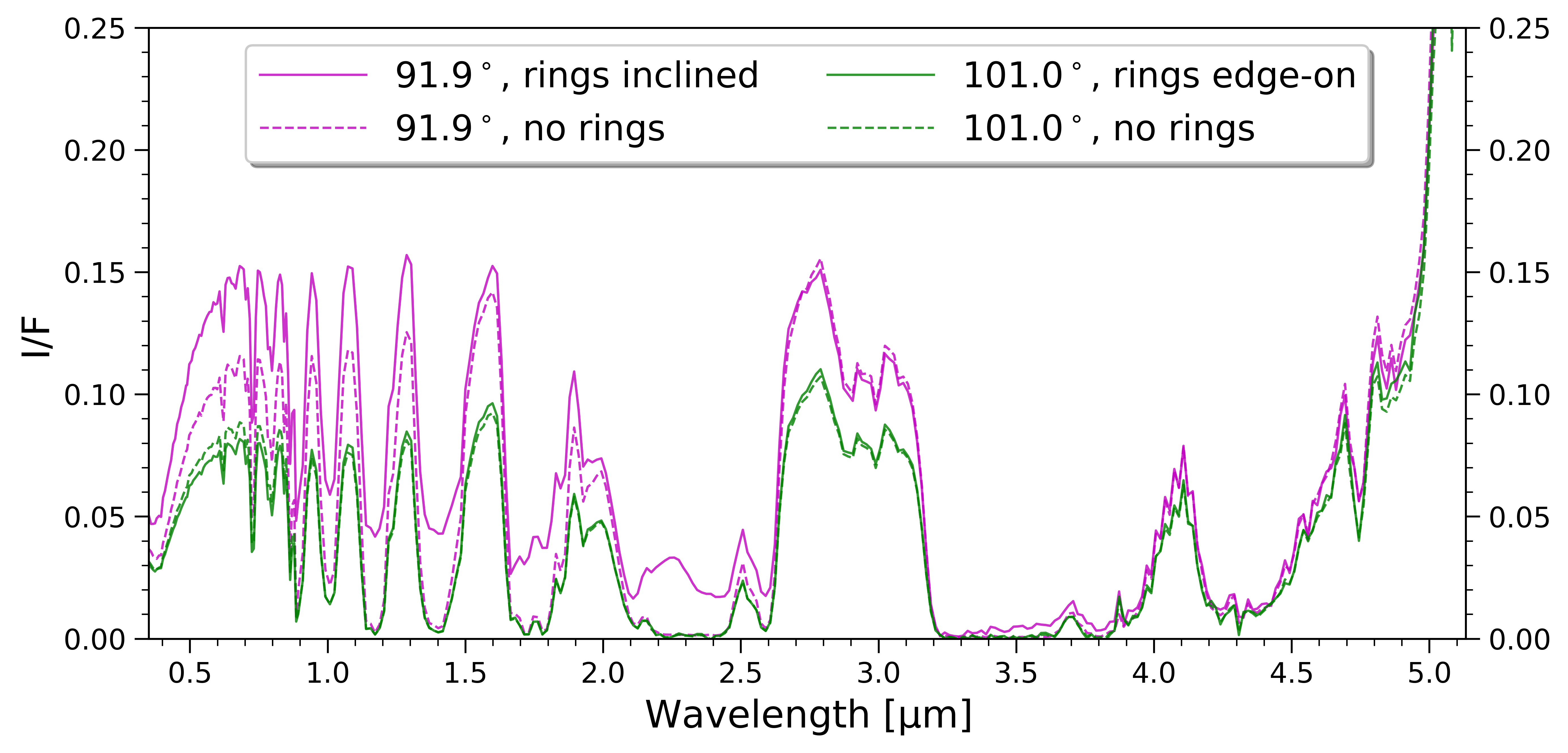}
\end{center}
\caption{Spectral comparison of two ring geometries: edge-on rings from cube V1518276136\_1 (green) and rings at an inclination of approximately $20^\circ$ from cube V1469259344\_1 (magenta). Spectra with the rings and their shadows removed are shown for comparison.}
\label{fig:rings}
\end{figure*}

Composed almost entirely of water ice, Saturn's rings are compositionally different than Saturn itself. Water has few absorption bands in the visual and near-infrared, giving the A Ring and B Ring end member spectra a relatively flat appearance in Figure \ref{fig:saturnem}. Therefore, the rings represent a minority share of the flux outside absorption bands, but are the contributing factor within them. This result illustrates the distinctive effect rings have on disk-integrated spectra and is expanded upon in the following section.

\subsection{Discussion}
\label{subsec:satdiscussion}

Saturn, with its impressive ring system, provides an interesting test case for our ability to infer the presence of icy rings from point-source spectra alone. During its time in the Saturn system, \emph{Cassini} observed Saturn from a range of orbital inclinations, from edge-on rings up to a $20^\circ$ inclination. A comparison of edge-on and inclined ring spectra is shown in Figure \ref{fig:rings}. With edge-on rings, the spectrum shows little variation from its ringless counterpart. The slight decrease in the visual and near-infrared is the consequence of the rings casting shadows on the disk, resulting in less reflected light. In contrast, the inclined ring spectrum varies significantly from its ringless counterpart in the visual and near-infrared. The most significant difference is the drastic percentage-wise increase within the major absorption bands. This result shows that the rings prevent the planet from appearing dark in the absorption bands, so the presence of water or ice can be readily observed in directly imaged brown dwarfs and exoplanets. Periodic temporal variations in spectra that oscillate between those shown in Figure \ref{fig:rings} would provide strong evidence for the presence of a significant ring system. \citet{Arnold_2004} and \citet{Dyudina_2005} showed that rings around exoplanets could be inferred from photometric variations, while this study shows that they can be found from spectral analysis. Used in tandem, these techniques could be a powerful tool for discovering exoplanet rings.

In addition, the nightside end member spectra have minor features that resemble those in the daytime spectra (see the lower pane in Figure \ref{fig:saturnem}). We conclude that these features are evidence of the nightside being illuminated by ringshine. Such features are not present in the nightside end member spectra of Jupiter, further supporting this conclusion. While this phenomenon cannot be directly observed in exoplanets or brown dwarfs, it should be accounted for when using these spectra for comparison to models.

\section{Conclusions}
\label{sec:conclusions}
Extrasolar gas giants and brown dwarfs are known to exhibit photometric and spectroscopic variability, with the most significant variations primarily in the infrared, the result of patchy clouds. Modeling the composition and evolution of these clouds is a primary research objective in the brown dwarf and exoplanet communities, but spatially resolvable spectra are needed to verify current and future models. Such data is surprisingly sparse, so this study aims to fill that data gap and builds upon previous work in four ways:
\begin{enumerate}[label=(\roman*)]
\item We present spectra of Saturn in addition to Jupiter, providing an additional analog object
\item Our work has much finer spectral resolution than previous studies
\item We present end member spectra in addition to disk-integrated spectra
\item Our data are publicly available for use in future studies, specifically for training atmosphere models
\end{enumerate}

The ground truth spectral data of Jupiter and Saturn provided in this study will act as spatially resolved comparisons to distant directly imaged brown dwarfs and extrasolar gas giants. This repository of empirical data will help verify current and future atmospheric models, as well as provide a baseline for novel instruments such as the Planet as Exoplanet Analog Spectrograph (PEAS) described in \citet{Martin_2020}. Disk-integrated spectra are analogous to current direct imaging spectra and those we will obtain from future missions, while the end member spectra provide spatial context. 

Another interesting result is the potential to infer the presence of icy rings solely from temporal variations in point source spectra of extrasolar gas giants. As the purpose of this study is to provide much needed empirical disk-integrated spectra of the solar system gas giants and not to assess exo-ring detection methods, the utility of this result was not explored further. However, with the potential this technique has, especially when used alongside photometric exo-ring detection methods, further investigation is certainly merited.

D. C., J. B., and J. F. are all supported by the NASA \emph{Cassini} Data Analysis Program, grant number 80NSSC19K0897.

\bibliography{references.bib}{}
\bibliographystyle{aasjournal}

\end{document}